# Focused ion beam direct writing of magnetic patterns with controlled structural and magnetic properties

Michal Urbánek[1,2], Lukáš Flajšman[1], Viola Křižáková[2], Jonáš Gloss[3], Michal Horký[1], Michael Schmid[3] and Peter Varga[1,3]

[1]*CEITEC BUT, Brno University of Technology, Purkyňova 123, 612 00 Brno, Czech Republic*
[2]*Institute of Physical Engineering, Brno University of Technology, Technická 2, 616 69 Brno, Czech Republic*
[3]*Institute of Applied Physics, TU Wien, 1040 Vienna, Austria*
Contact address: michal.urbanek@ceitec.vutbr.cz

**Focused ion beam irradiation of metastable $Fe_{78}Ni_{22}$ thin films grown on Cu(100) substrates is used to create ferromagnetic, body-centered-cubic patterns embedded into paramagnetic, face-centered-cubic surrounding. The structural and magnetic phase transformation can be controlled by varying parameters of the transforming gallium ion beam. The focused ion beam parameters as ion dose, number of scans, and scanning direction can be used not only to control a degree of transformation, but also to change the otherwise four-fold in-plane magnetic anisotropy into the uniaxial anisotropy along specific crystallographic direction. This change is associated with a preferred growth of specific crystallographic domains. The possibility to create magnetic patterns with continuous magnetization transitions and at the same time to create patterns with periodical changes in magnetic anisotropy makes this system an ideal candidate for rapid prototyping of a large variety of nanostructured samples. Namely spin-wave waveguides and magnonic crystals can be easily combined into complex devices in a single fabrication step.**

Direct writing of magnetic patterns by focused ion beam (FIB) irradiation[1] presents a favorable alternative to the conventional lithography approaches. It removes the need for further processing of the specimen and allows for a rapid prototyping of a large variety of nanostructured samples. Since the pioneering work of Chappert et al.[2], many different approaches to ion-beam-induced magnetic patterning have been studied, including modification of magnetic anisotropies[2], coercivity, exchange bias[3] or the magnetization of the material[4]. However, these approaches often lead to ferromagnetic structures embedded in a ferromagnetic (or antiferromagnetic) matrix. For many applications, it is more suitable to have ferromagnetic elements surrounded by nonmagnetic regions. This can be done either by destruction of magnetism in multilayers by e.g. ion-induced alloying[5] or via a positive process by creating ferromagnetic elements by ion-induced change of chemical[6,7] or structural order[8]. Another possible approach is to use ion-induced chemical reactions to create magnetic patterns[9-11].

Metastable face-centered cubic (fcc) Fe thin films[12,13] are good candidates for magnetic patterning, because they are paramagnetic at room temperature and can be transformed by ion-beam irradiation to ferromagnetic body-centered cubic (bcc) Fe[8]. Unfortunately, there is a thickness limit as fcc Fe films thicker than approx. 2 nm transform spontaneously to bcc[12]. It is possible to overcome this thickness limit by stabilizing the fcc phase either by depositing the Fe at increased CO background pressure[14,15] or by alloying with Ni[16]. In this work we use 8-nm-thick Ni-stabilized fcc Fe films as nonmagnetic template and study the influence of FIB parameters on the structural and magnetic properties of transformed patterns. We show that it is possible to control not only the degree of transformation (saturation magnetization) but also the growth of specific crystallographic domains exhibiting different magnetic anisotropies (uniaxial anisotropy directions).

The films were grown in an ultra-high vacuum (UHV) system by evaporation from $Fe_{78}Ni_{22}$ (2 mm thick rod, purity 99.99%) heated by electron bombardment. Prior to the experiments, the Cu(100) crystals were cleaned by several cycles of sputtering (2 keV $Ar^+$ ions, 30 min) and annealing (600 °C, 10 min). The cleanliness of the surface as well as the film composition was checked by Auger Electron Spectroscopy (AES). The pressure during the deposition was $5\times10^{-10}$ mbar and the deposition rate 0.02 Å/s (calibrated by a quartz-crystal microbalance) resulted in a deposition time of approx. 1 h for 8-nm films. To suppress high-energy ions which may modify the growth mode of the films[17], a repelling voltage of +1.5 kV was applied to a cylindrical electrode in the orifice of the evaporator. After the deposition, the crystallographic structure of the films was checked by low-energy electron diffraction (LEED).

The Cu crystal with deposited metastable film was then removed from UHV and transferred into the high vacuum chamber of the scanning electron microscope equipped with focused ion beam column (FIB-SEM, Lyra3, Tescan) where we conducted the FIB transformation. The residual pressure in the FIB-SEM vacuum chamber during transformation was $9\times10^{-7}$ mbar. For the experiments we used the following nominal parameters of the gallium ion beam: acceleration voltage 30 kV, beam current 145 pA, beam spot size 30 nm and scanning step size 10 nm. First, we performed a dose test where we transformed rectangles (6 μm ×14 μm) with an increasing ion dose. The transformation was performed by two different approaches: 1) by performing 100 fast scans over the full area of the rectangle and 2) by applying the full ion dose in one (slower) scan. The total irradiation time was the same in both cases. After the transformation, we imaged the transformed areas by SEM. Although the sample surface after transformation was perfectly flat, by using an electron energy of 5 keV and a conventional Everhart-Thornley (SE) detector we were able to observe a clear contrast between irradiated and non-irradiated areas. The contrast was reversing from dark to white and back upon tilting the sample ±10° from the normal and also upon rotation (with 6-fold symmetry in bcc areas and 8-fold symmetry in fcc areas), which points to its crystallographic origin[18]. This crystallographic contrast cannot be fully quantified, but it is sufficient to image the difference between untransformed fcc Fe and transformed bcc Fe areas and also to distinguish different orientations of bcc domains after the transformation. Additionally, we measured the Kerr ellipticity (which is proportional to magnetization) of the transformed areas with our home built micro-Kerr magnetometer[19].

The results of the dose test are shown in FIG. 1. The graph in FIG 1. a) shows dependence of the Kerr ellipticity (degree of transformation) on the ion dose for different number of scans. Each point in the graph is from a separate experiment. The results show clear difference in the transformation process when the structures are transformed by using either multiple scans over the same area (dashed line with open circles) or by single scan only (solid line with open triangles). When irradiating the material by multiple passes of the ion beam the magnetization of the structures increases linearly from the background value of $5\times10^{-6}$ mrad to the value of $0.9\times10^{-4}$ mrad at an ion dose of $2\times10^{15}$ ions/cm$^2$. This suggests a stochastic process of transformation resulting in small bcc nuclei, where the number of nuclei is proportional to the number of incident ions (keeping the probability of creating the bcc nuclei by incident ion constant). From the linear fit of the Kerr ellipticity and assuming maximum measured Kerr ellipticity equals to fully transformed layer we can estimate the transformation efficiency of approx. 3 Fe or Ni atoms per incoming $Ga^+$ ion [see inset in FIG. 1 a)]. After the saturation ion dose (maximal magnetic signal) is reached the ion-beam-induced intermixing and sputtering processes lower the magnetization down to the point where all the iron and nickel has been sputtered off and no magnetic signal is observed anymore.

In the case of a single scan the magnetization in the low-dose regime also increases linearly (suggesting the same mechanism as in multi-scan approach) yet when a critical ion dose of $3\times10^{15}$ ions/$cm^2$ is reached the transformation efficiency suddenly increases to approx. 12 Fe or Ni atoms per incoming $Ga^+$ ion. The ion beam is now irradiating the fcc-bcc boundary with sufficient ion flux to achieve steady state migration of the bcc structure into the fcc surroundings[20,21].

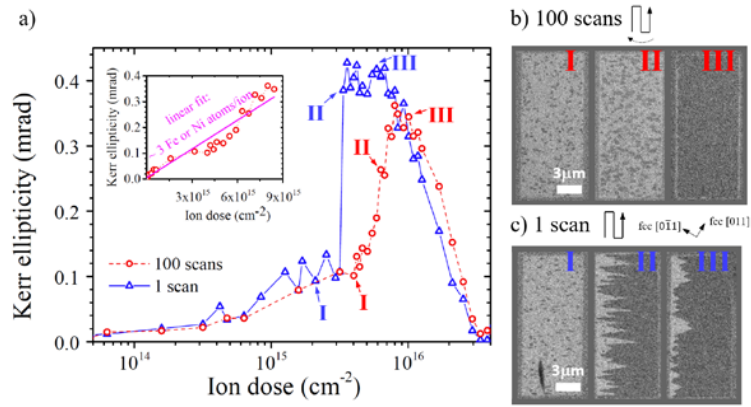

*FIG. 1. Dose test and development of FIB induced transformation. a) Plot of the dependence of the Kerr ellipticity on the ion dose for different number of scans. Inset shows linear fit of the Kerr ellipticity with the constant transformation efficiency of approx. 3 Fe or Ni atoms per incoming $Ga^+$ ion. b) Evolution of the transformation when the rectangles were irradiated by multiple scans over the whole rectangle area. c) Evolution of the transformation when the rectangles were irradiated by single scan. The numbers I, II and III represent corresponding points on the curves shown in plot a).*

The two different regimes of transformation are also illustrated in SEM images [FIG. 1 b) and c)]. The crystallographic contrast[18] allows us to clearly distinguish three gray levels: the intermediate gray level corresponding to the untransformed areas (observable around the rectangular patterns), bright corresponding to areas in the early stage of transformation and dark indicating the fully transformed bcc Fe thin film. For low doses the transformed areas look the same for single-scan and multiple-scan transformations. The irradiated rectangles I consist of a bright area with several darker, fully transformed spots. For the rectangles II, irradiated with intermediate ion doses, the situation is different. For a single scan (scanning vertical lines from the bottom left corner), it is evident that once the transformation starts it is propagating in expanding triangles from individual nuclei until the triangles connect and the film is fully transformed. The rectangle irradiated by multiple scans looks differently. The individual nuclei are becoming larger and denser but they are homogeneously distributed over the whole irradiated area. The image of the rectangle irradiated with the single scan and a high dose [FIG. 1 c), rectangle III] shows a small but still present partially transformed bright area on the left side while the rest of the rectangle is fully transformed. The image of the rectangle irradiated with the multiple scans and a high dose [FIG. 1 b), rectangle III] shows fully transformed area but the resulting magnetization measured inside the rectangle is lower than in the dark area of FIG. 1 c), rectangle III [compare also with the graph in FIG. 1 a)]. This is because the maximum is reached at a later stage, where sputtering and intermixing already decreases the magnetization. These results show that single-scan transformation is much more efficient than multiple-scan transformation and once the initial bcc nuclei are formed, then the transformation proceeds mainly via grain growth of the already transformed areas. Once the initial grain is transformed it is easier to move the grain boundary via collision induced migration of vacancies and interstitials at the boundary[20,21].

In the second experiment we transformed a 15x15 $\mu m^2$ square in a single scan with an ion dose of $2\times10^{15}$ ions/cm$^2$. The ion beam was scanning in square spiral from the center of the rectangle towards the border. The resulting SEM image [FIG. 2. a)] shows clear division of the rectangle into four triangular domains; each domain corresponds to a different scanning direction. Inside of each triangular domain we can observe an additional texture. Unfortunately, the SEM observation does not allow to extract any quantitative information about the crystallography of the areas with different contrast[18].

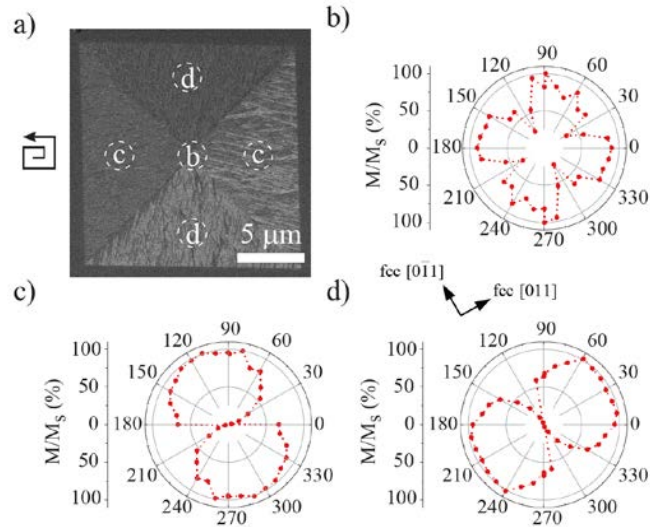

*FIG. 2. Rectangle transformed by spiral scanning. a) Crystallographic contrast in SEM shows division into four domains resulting from different FIB scanning directions. Polar plots of remanent magnetization measured by micro-Kerr magnetometry show four-fold magnetic anisotropy in the center of the rectangle (point b) and uniaxial anisotropies with different directions inside the crystallographic domains (points c and d).*

Magnetic measurements provide further insight into the behavior of the material. We used the micro-Kerr magnetometer[19] to measure the angular dependence of the remanent magnetization in the center of the transformed square and in the center of each triangular domain. The spot size of the micro-Kerr magnetometer was approx. 1 μm. In the center of the square the plot of remanent magnetization shows clear four-fold magnetic anisotropy [FIG. 2. b)], whereas inside the triangular domains the magnetic anisotropy is clearly uniaxial. Moreover, the direction of the uniaxial anisotropy changes with the FIB scanning direction [FIG. 2. c), d)].

To study the dependence of the magnetic anisotropy direction on the scanning direction we transformed 36 circles with 10 μm diameter. The circles were transformed by linear scanning with varying angle of FIB scanning starting from fcc [011] direction in 10° steps. The results of the experiment are shown in FIG. 3 a). When the direction of FIB scanning was between 0° and 90° (fcc [011] and fcc [0$\bar{1}$1] directions) the resulting magnetic anisotropy direction, represented by the direction of the easy axis rotated by approx. 20° between 35° and 55°. When the direction of FIB scanning passed 90° (fcc [0$\bar{1}$1] direction) the resulting easy axis direction jumped from 55° to 125°. With further increase of the scan angle, the resulting easy axis direction further gradually changed from 125° to 145°. At 180° (fcc [0$\bar{1}\bar{1}$]) the easy axis again jumped from 145° to 35° and the angular dependence continued symmetrically in third and fourth quadrant, with continuous rotation for FIB scanning in between the fcc low-index directions and jumps when the FIB scanning direction passed the fcc low-index directions.

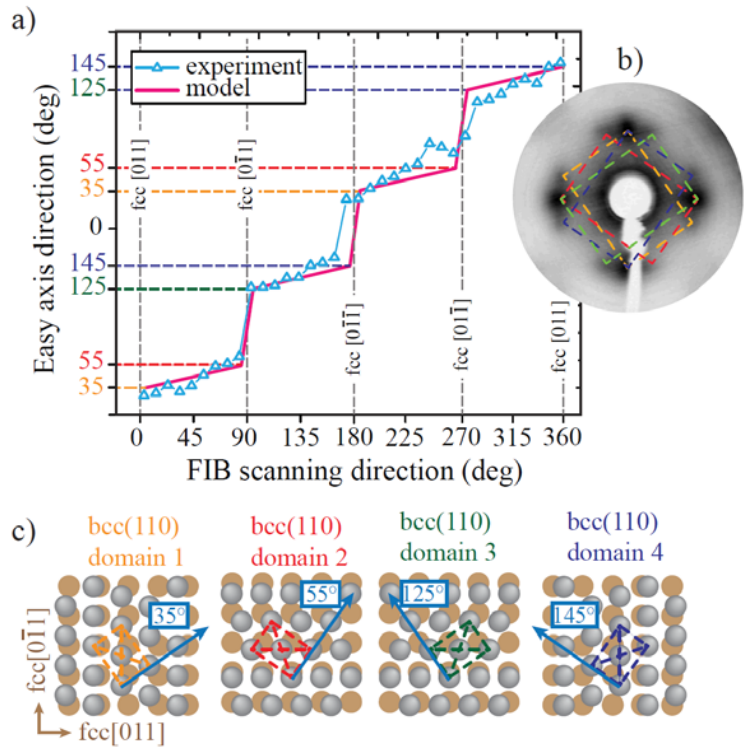

FIG. 3. a) Magnetic anisotropy (easy axis direction) as a function of the FIB scanning direction (0° corresponds to fcc [011]). b) LEED pattern showing four bcc(110) domains after transformation by a broad ion beam. c) Schematic of all four bcc(domains) with arrows indicating directions of the easy axes.

To explain the behavior of the evolution of the magnetic uniaxial anisotropy with respect to the direction of the FIB scanning we need to look at the crystallographic structure of the bcc Fe thin films on Cu(100). The LEED pattern [see FIG. 3 b)] of the 8 nm thin film transformed in UHV by broad $Ar^+$ ion beam shows four possible bcc(110) domains formed in a Pitsch orientational relationship[22]. The magnetic easy axis for bcc Fe is aligned with its <001> directions[23]. FIG. 3 c) shows of all four possible bcc(110) domains, with blue arrows indicating the angles of the easy axes. For these domains, the azimutal angles of the easy axes are 35°, 55°, 125° and 145°.

Putting together the magnetic and structural data reveals the behavior of the FIB-induced transformation. In case of transformation by a broad ion beam or by isotropic scanning by FIB (and also when using multiple FIB scans) the transformed film contains all four bcc(110) domains [see FIG. 3. b)] and exhibits four-fold magnetic anisotropy [see FIG. 2. b)]. The linear single-scan FIB transformation results in uniaxial magnetic anisotropy [see FIG. 2 c), d)] and the direction of the anisotropy depends on the direction of FIB scanning [see FIG. 3 a)]. The experimental data fit to the model where FIB scanning in between fcc low-index directions preferentially forms bcc domains which have [001] direction parallel, or close to parallel to FIB scanning direction. For example, when the FIB is scanning between 0° and 90° then there is preferential nucleation of the domains with [001] directions at 35° and 55° [domain 1 and domain 2 in FIG. 3. c)]. The FIB scanning angle can control the ratio of transformed domains and the easy axis direction can be continuously rotated between 35° and 55°. When the FIB scanning angle exceeds 90° (fcc [0$\bar{1}$1] direction), then the other two bcc(110) domains are preferred and the magnetic easy axis jumps by 70° from 55° to 125°. Then, by further increase of the FIB scanning angle from 90° to 180° it is again possible to control the ratio of transformed domains [domain 3 and domain 4 in FIG. 3. c)] and to rotate the easy axis between 125° and 145°. The exact reason why the direction of FIB scanning can control the nucleation of individual bcc domains is not clear. The most probable explanation is uniaxial strain propagating perpendicularly to the FIB scanning direction.

The films described in this paper are well suited to prepare magnetic patterns or structures which are extremely difficult or impossible to prepare by conventional lithography techniques. In FIG. 4 a) we show a magnonic crystal consisting of 500 nm wide stripes with alternating magnetic anisotropy. In FIG. 4 b) is another magnonic crystal with modulated magnetization. The modulation in magnetization can be either in steps, or it is also possible to fabricate a gradual magnetization transition. All these structures are results of pure magnetic patterning without any apparent topography on the irradiated structures. The contrast in SEM images is purely crystallographic.

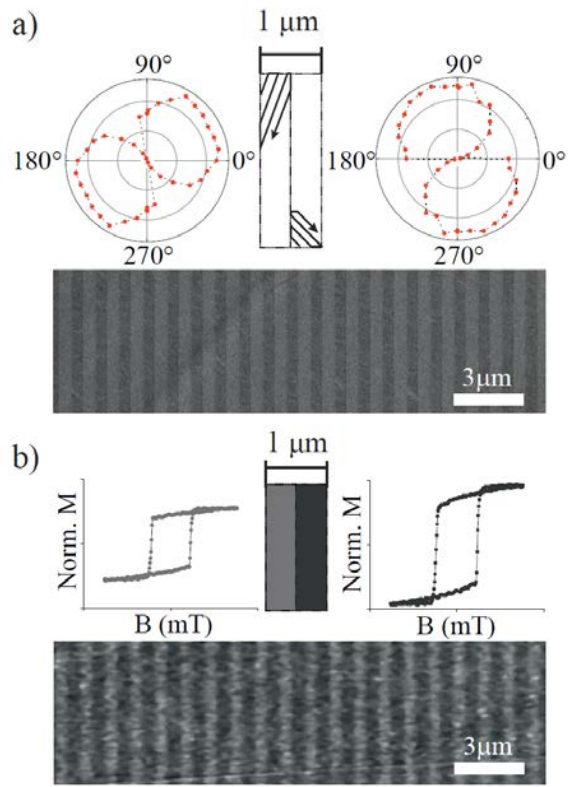

*FIG. 4. Examples of magnetic patterns. a) Magnonic crystal with periodical changes in magnetic anisotropy. b) Magnonic crystal with modulated magnetization.*

In summary, we have presented very powerful method for magnetic pattering by direct FIB writing. The system allows precise control of the magnetic parameters of the transformed areas. We have shown that it is possible to control degree of transformation (magnetization) by selecting proper ion dose and using multiple scans over the sample area. Even more important, we have shown that it is also possible to control the magnetic anisotropy of the transformed patterns by changing the FIB scanning direction. With linear scanning, the bcc(110) domains having [001] directions (easy axes) parallel, or close to parallel to the FIB scanning direction are preferentially formed. The examples of transformed patterns with sub 100-nm transitions show that FIB patterned metastable $Fe_{78}Ni_{22}$ thin films on Cu(100) can be used as rapid prototyping platform for many spintronic and magnonic applications.

This research has been financially supported by the joint project of Grant Agency of the Czech Republic (Project No. 15-34632L) and Austrian Science Fund (Project No. I 1937-N20). The FIB transformation was carried out in CEITEC Nano Research Infrastructure (ID LM2015041, MEYS CR, 2016–2019). L.F. was supported by Brno PhD talent scholarship.